# Analysis of vessel traffic flow characteristics in inland restricted waterways using multi-source data

Wenzhang Yang[1], Peng Liao[1]*, Shangkun Jiang[2], Hao Wang[1]

[1] *School of Transportation, Southeast University, Nanjing, Jiangsu, 211189, China*
[2] *Department of Civil and Coastal Engineering, University of Florida, Gainesville, FL, 32611, USA*

**ABSTRACT**

To effectively manage vessel traffic and alleviate congestion on busy inland waterways, a comprehensive understanding of vessel traffic flow characteristics is crucial. However, limited data availability has resulted in minimal research on the traffic flow characteristics of inland waterway vessels. This study addresses this gap by conducting vessel-following experiments and fixed-point video monitoring in inland waterways, collecting multi-source data to analyze vessel traffic flow characteristics. First, the analysis of vessel speed distribution identifies the economic speed for vessels operating in these environments. Next, the relationship between microscopic vessel speed and gap distance is examined, with the logarithmic model emerging as the most accurate among various tested models. Additionally, the study explores the relationships among macroscopic speed, density, and flow rate, proposing a novel piecewise fundamental diagram model to describe these relationships. Lastly, the inland vessel traffic states are categorized using K-means clustering algorithm and applied to vessel navigation services. These findings provide valuable insights for enhancing inland waterway transportation and advancing the development of an integrated waterway transportation system.

**Keywords**：Inland waterway; inland vessel traffic; traffic flow characteristic; fundamental diagram model; traffic state recognition.

**Highlights**

✓ Vessel traffic flow characteristics in inland restricted waterways are analyzed using multi-source data.

✓ The relationship between microscopic vessel speed and gap distance is examined.

✓ A piecewise fundamental diagram model is proposed to describe the relationships among macroscopic speed, density, and flow rate.

✓ Vessel traffic states are categorized and applied to vessel navigation services.

## 1. Introduction

Inland waterway transportation is a crucial element of the integrated transportation system, significantly contributing to the optimization of industrial layouts and the promotion of economic growth in river basins (Nayak et al., 2024; Oztanriseven and Nachtmann, 2020; Welch et al., 2022). Recently, there has been a surge in demand for inland waterway transportation. In China, the freight volume for inland waterways reached 4.791 billion tons in 2023, surpassing ocean freight (4.577 billion tons) and approaching railway freight (5.035 billion tons). This high freight volume has created a gap between waterway capacity and transportation demand, leading to increased delays in vessel traffic (Kang et al., 2022; Tan et al., 2015; Yan et al., 2017). To mitigate delays under congested conditions, it is essential to implement vessel traffic control and develop effective waterway traffic management strategies, all of which require a thorough understanding of the fundamental characteristics of vessel traffic flow in inland waterways.

---

* Corresponding author. Tel.: +86 25 5209 1265, *E-mail address*: pliao@seu.edu.cn.

Traffic flow originally referred to the continuous movement of vehicles on the road, exhibiting fluid-like characteristics due to their ongoing motion (Chen et al., 2024a&b; Hua et al., 2011; Jin et al., 2013; Yang et al., 2023b&c). More broadly, it also encompasses traffic involving other carriers and pedestrians, such as vessel traffic in inland or maritime waterways. Early studies on vessel traffic flow emerged from maritime navigation during the international shipping boom of the 1950s (Toyoda and Fujii, 1971; Wepster, 1978). Like vehicle traffic flow, vessel traffic theory focuses on macro and micro characteristics, using statistical analysis methods (Wen et al., 2015; Li et al., 2015; Liu et al., 2017b; Liu et al., 2020). For example, in macroscopic traffic flow theory, key parameters include flow rate ($q$), average speed ($v$), and density ($k$), with flow rate being the product of speed and density. Microscopic traffic flow theory, however, examines detailed vessel behaviors, such as vessel-following actions. Building on these theories, various studies have explored topics such as the maximum capacity of waterways (Liu et al., 2020), the relationship between vessel speed and density (Huang et al., 2019; Kang et al., 2018; Liao et al., 2022), and the prediction of vessel traffic trends (Dong et al., 2024; Li et al., 2023b; Xing et al., 2023).

As an applied discipline, traffic flow theory relies heavily on real data collection. Recently, with the widespread use of the Automatic Identification System (AIS), there is now ample observational data available to study maritime traffic flow characteristics (Kang et al., 2022; Yu et al., 2020). For instance, Kang et al. (2018) and Huang et al. (2019) examined the relationship between basic parameters of vessel traffic flow using AIS data and showed that, despite being influenced by complex environmental factors like tides, currents, waves, and winds, vessel traffic flow shares many similarities with vehicle traffic flow. Additionally, some studies have demonstrated that integrating AIS with machine learning can address more complex issues in vessel traffic flow (Li and Yang, 2023; Li et al., 2023a; Yang et al., 2024).

The studies mentioned above primarily focus on maritime vessel traffic issues, with only a few examining inland waterway vessel characteristics (Hart et al., 2023; Jin et al., 2023; Muthukumaran et al., 2022; Yang et al., 2023a). This gap is largely attributable to the limited availability of AIS data for inland waterways. In these areas, AIS base stations and signal coverage are often incomplete, and some vessels lack advanced AIS equipment. This results in incomplete data, making reliable analysis of vessel characteristics challenging. To address this, some research has involved conducting field experiments in inland waterways to gather empirical data. For instance, Yang et al. (2023a) carried out vessel-following experiments and applied various car-following models from vehicle traffic flow to analyze their data, finding that some models exhibited effective transferability. Additionally, Jin et al. (2023) and Hart et al. (2023) developed new vessel-following models based on their data. However, these studies mainly concentrate on the microscopic behavior of vessels and tend to overlook the macroscopic characteristics of vessel traffic flow in inland waterways. Moreover, the data in these studies is generally insufficient and typically derived from a single source.

This study aims to analyze the characteristics of vessel traffic flow in inland restricted waterways using a variety of data sources. These sources include video surveillance data from Jiangsu Province, China, as well as data from vessel-following field experiments. The objective is to provide a comprehensive analysis of vessel traffic flow from a novel perspective. The study will start by examining the data to understand the microscopic behaviors of vessels in these restricted waterways, identify the macroscopic features of vessel traffic flow, and show its applications for waterway transportation management. Ultimately, this research seeks to provide theoretical support for advancing inland waterway transportation and contribute to the development of a modern, integrated waterway transportation system.

The remainder of this paper is organized as follows: Section 2 introduces the data sources and outlines the preliminary data processing. Section 3 provides a comprehensive analysis of vessel traffic flow characteristics. In Section 4, fundamental diagram models are used to fit the data, and a new model for vessel traffic flow is proposed. Section 5 discusses traffic state recognition within vessel traffic flow and its applications. Finally, Section 6 presents the conclusions.

## 2. Data

This paper examines vessel traffic flow characteristics using two types of data: video surveillance data and vessel-following experiment data. The video surveillance data is sourced from the perception channel information technology on the Grand Canal in the Wuxi section. This dataset primarily features single vessels with relatively large distances between them, indicating a tendency towards free-flow conditions. In contrast, the vessel-following experiment data consists of continuous observations of the same fleet, which is more representative of non-free-flow conditions.

### 2.1. Video surveillance data

The video surveillance data captured the navigation details of the Wuxi section of the Grand Canal. This section of the canal is approximately 90 meters wide and has a navigable depth of about 4 meters, which accommodates cargo vessels up to 1,000 tons. The Wuxi section is equipped with an advanced perception system designed to comprehensively monitor and record vessel navigation data. This system integrates vessel traffic flow detection technologies, combining video surveillance, AIS, and radar detection (Chen et al., 2020; Park et al., 2022). As illustrated in Fig. 1, the system captures video footage and timestamps of vessels as they navigate through the canal, recording real-time parameters such as instantaneous speed, vessel size, tonnage, load capacity, type, and direction of travel.

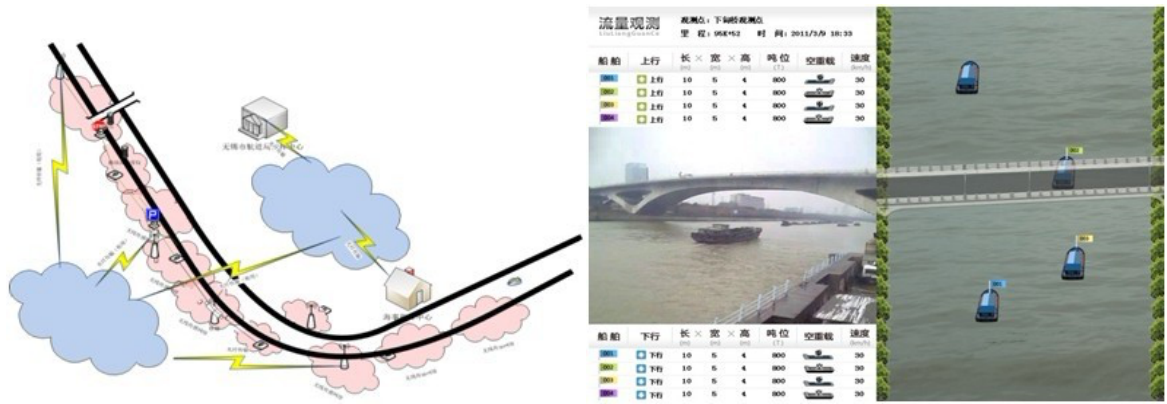

**Fig.1.** The perception channel of the Grand Canal in the Wuxi section.

A total of 74,496 vessel navigation data points were collected from the Wuxi section of the Grand Canal during January, May, August, and October 2014, as shown in Table 1. This dataset comprises 74,019 observations of individual vessels and 477 observations of fleets. Analysis of the data reveals that single vessel observations represent over 99% of the total, whereas fleet observations account for less than 1%. This distribution suggests that the vessel traffic flow in the canal is predominantly comprised of individual vessels.

**Table 1**
Classification and summary of vessels of the Grand Canal in the Wuxi section.

|  |  | January | | May | | August | | October | |
|---|---|---|---|---|---|---|---|---|---|
|  |  | Single | Fleet | Single | Fleet | Single | Fleet | Single | Fleet |
| Upstream vessel | Number | 12286 | 87 | 13504 | 75 | 4692 | 29 | 12755 | 64 |
|  | Proportion | 99.30% | 0.70% | 99.45% | 0.55% | 99.39% | 0.61% | 99.50% | 0.50% |
| Downstream vessel | Number | 8051 | 82 | 12231 | 74 | \ | \ | 10500 | 66 |
|  | Proportion | 98.99% | 1.01% | 99.40% | 0.60% | \ | \ | 99.38% | 0.62% |
| Total number |  | 20337 | 169 | 25735 | 149 | 4692 | 29 | 23255 | 130 |
| Proportion |  | 99.18% | 0.82% | 99.42% | 0.58% | 99.39% | 0.61% | 113.41% | 0.63% |

Fig. 2 illustrates the distribution of vessel sizes in the Wuxi section of the Grand Canal. The data reveals that most vessels have lengths ranging from 40 meters to 50 meters, accounting for 62.28% of the total. The next most common length range is 30 meters to 40 meters, comprising 32.03%. Vessels longer than 50 meters or shorter than 30 meters are less common, representing only 5.69% of the total. Additionally, the majority of vessels have widths between 6 meters and 9 meters. Given the similarity in length and width dimensions, these vessels are well-suited for analyzing and researching the traffic characteristics of inland waterways.

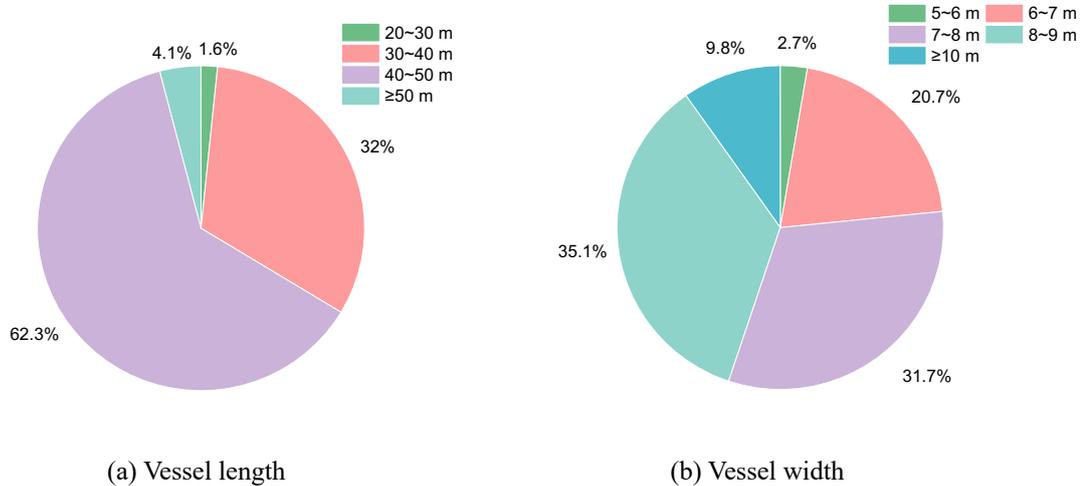

(a) Vessel length      (b) Vessel width

**Fig.2.** The size distribution of vessels in the Wuxi section of the Grand Canal.

Based on video surveillance data, the flow rate is determined by counting the number of vessels passing through a specified cross-section per unit of time. Vessel speed is then calculated as the spatial average speed of all vessels within the corresponding time segment. With the flow rate and speed data in hand, vessel density can be derived using the three-parameter relationship of traffic flow. This methodology provides comprehensive data for all three key parameters of vessel traffic flow.

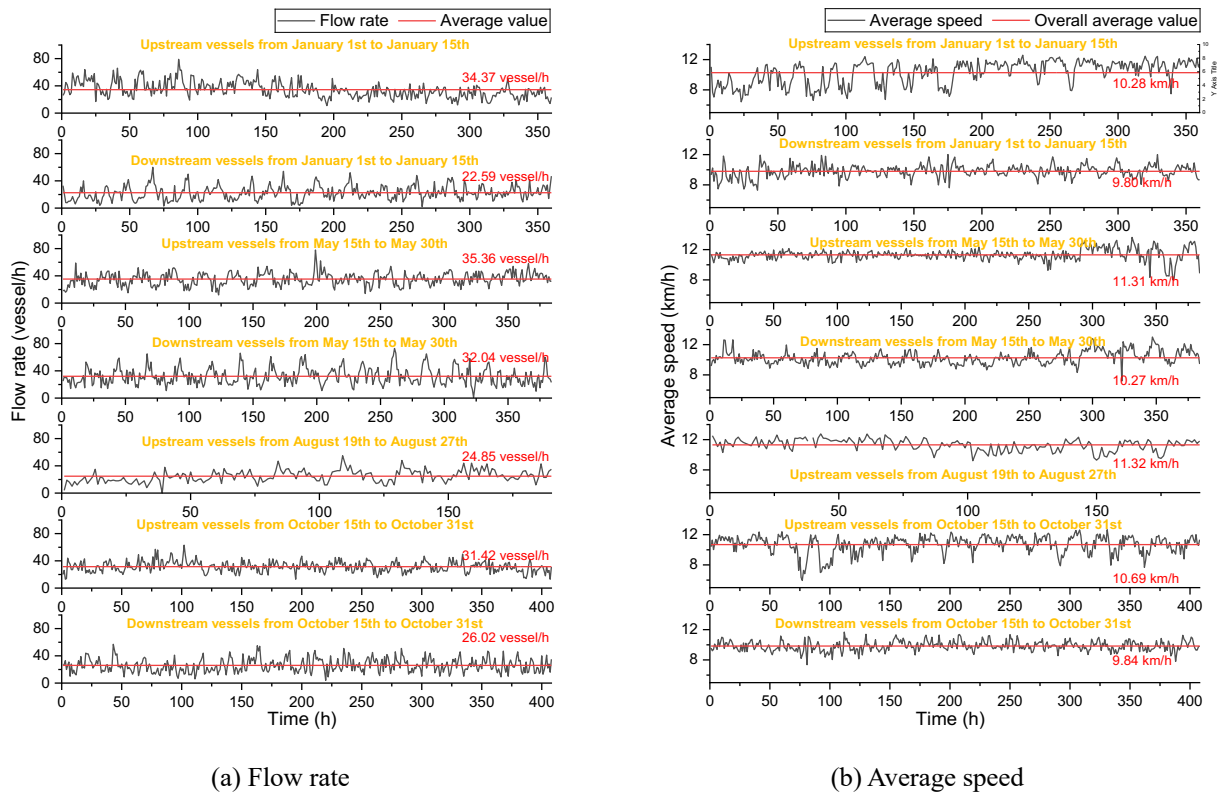

(a) Flow rate      (b) Average speed

**Fig.3.** Flow rate and average speed variation curves for the Grand Canal in the Wuxi section during January, May, August, and October of 2014.

Fig. 3 illustrates the variations in vessel traffic flow rates and average speeds in the Wuxi section of the Grand Canal for January, May, August, and October 2014. The average flow rate across all data points is 29.89 vessels per hour, and the overall average speed is 10.44 km/h. Generally, the upstream flow rate exceeds the downstream rate, and upstream vessels have a higher average speed compared to downstream vessels. This can be attributed to the higher proportion of loaded vessels traveling downstream. Specifically, only 26.72% of upstream vessels are loaded, whereas 88.63% of downstream vessels are loaded. Overall, vessel flow rates and average speeds in this channel are relatively stable, with no significant diurnal fluctuations. This data effectively represents the characteristics of inland waterway vessel traffic under free-flow conditions.

## 2.2. Vessel-following experiment data

To collect and analyze data under non-free-flow traffic conditions, we conducted three vessel-following experiments between 2019 and 2021, designated as Experiment 1, Experiment 2, and Experiment 3 (Jin et al., 2023; Yang et al., 2023a). The navigation segments and vessel directions are illustrated in Fig. 4, with additional background parameters provided in Table 2. Aerial imagery and a GNSS positioning schematic from these experiments are displayed in Fig. 5.

Each experiment involved eight vessels, with the first vessel acting as the leading vessel. In Experiments 1 and 3, the leading vessel was a smaller speedboat, while the remaining vessels were standard cargo vessels. Experiment 2 comprised eight cargo vessels. Table 3 lists the dimensions of the vessels used. All cargo vessels were similar in performance and size, with an average carrying capacity of 800 tons. The cargo vessels were loaded during Experiments 1 and 2, and empty during Experiment 3.

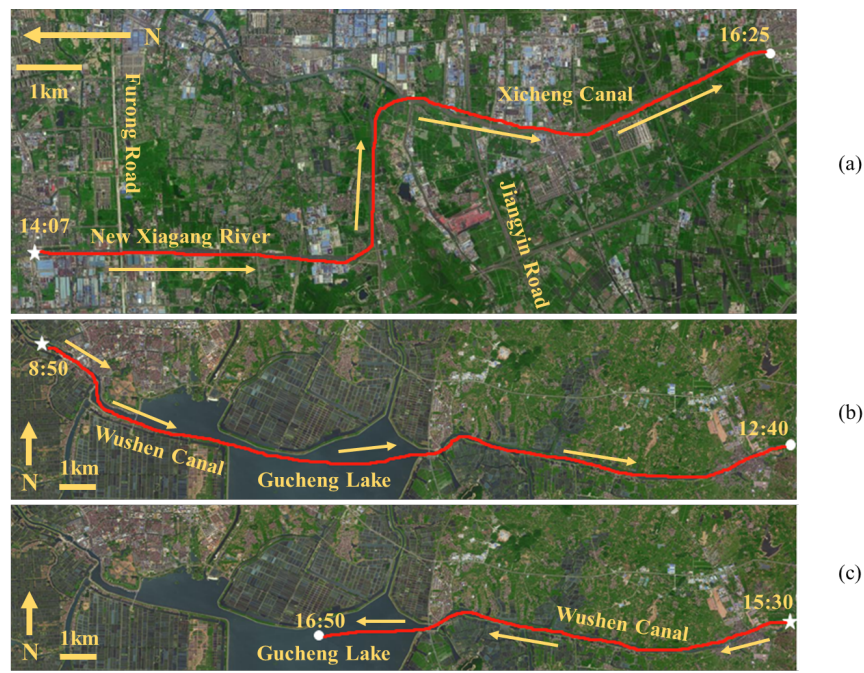

**Fig.4.** Waterway map of (a) Experiment 1, (b) Experiment 2, (c) Experiment 3. (The arrows represent the navigation direction.)

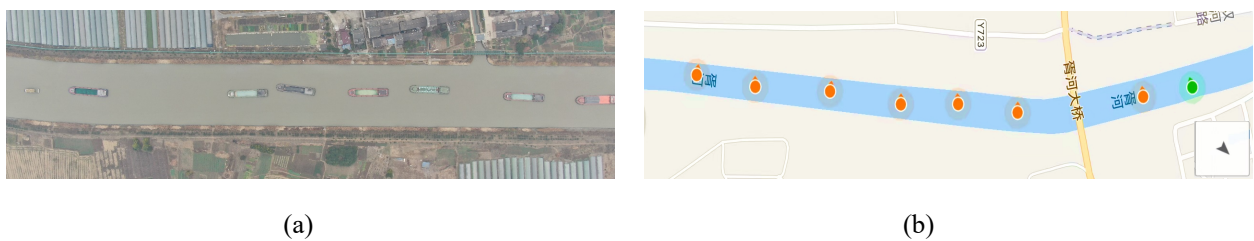

(a)          (b)

**Fig.5.** (a) Aerial photo of the fleet in Experiment 1, and (b) GNSS location of the fleet in Experiments 3.

**Table 2**
Background parameters of the three vessel-following experiments.

|  | Experiment 1 | Experiment 2 | Experiment 3 |
|---|---|---|---|
| Data | December 20, 2019 | October 24, 2021 | October 24, 2021 |
| Duration | 14:07~16:25 | 8:50~12:40 | 15:30~16:50 |
| Experimental location | Xicheng Canal | Wushen Canal | Wushen Canal |
| Navigation distance (km) | 20 | 28 | 13 |
| Channel width (m) | 70 | 80 | 80 |
| Wind speed (m/s) | 1 | 0.5 | 0.5 |
| Water flow speed (m/s) | 0.05 | 0.05 | 0.05 |
| Water flow direction | From south to north | From west to east | From west to east |
| Water depth (m) | 3 | 4.5 | 4.5 |
| Vessel loading condition | Loaded | Loaded | Empty |

**Table 3**
The length of vessels of three experiments (m).

| Experiment | No.1 | No.2 | No.3 | No.4 | No.5 | No.6 | No.7 | No.8 |
|---|---|---|---|---|---|---|---|---|
| 1 | 17.0 | 44.8 | 44.9 | 43.5 | 44.8 | 43.0 | 44.9 | 49.8 |
| 2 | 52.8 | 52.8 | 39.8 | 44.0 | 49.6 | 45.6 | 48.0 | 44.8 |
| 3 | 15.6 | 49.8 | 44.3 | 40.0 | 42.8 | 54.8 | 49.0 | 54.8 |

The vessel navigators were recruited on-site on the day of the experiment. During the experiments, the speed of the leading vessel was controlled according to the experimental plan. The other navigators were informed of and complied with the rule against overtaking the preceding vessel. The navigators had varying levels of experience, ranging from 5 to 40 years.

The positions of the vessels in these experiments were recorded every second by high-precision GNSS locators, with measurement errors within ±1 meter. Fig. 6 illustrates the speed and gap distance of the experimental vessels, calculated using the following formulas:

$$v_{i,j}(t) = \frac{\sqrt{(X_{i,j}(t+\Delta t) - X_{i,j}(t))^2 + (Y_{i,j}(t+\Delta t) - Y_{i,j}(t))^2}}{\Delta t} \, (i=1,2,3,...8; j=1,2,3.) \tag{1}$$

$$g_{i,j}(t) = \frac{\sqrt{(X_{i-1,j}(t) - X_{i,j}(t))^2 + (Y_{i-1,j}(t) - Y_{i,j}(t))^2}}{\Delta t} + d_{i-1,j} - d_{i,j} - L_{i-1,j} \, (i=2,3,...8; j=1,2,3.) \tag{2}$$

where $v$ and $g$ represent the speed and gap distance of the subject vessel, respectively; $L$ denotes the length of subject vessel; $d$ is the distance from the GNSS locator to the vessel's bow; $j$ denotes the number of experiments; $i$ means the number of the subject vessel in an experiment; $t$ means the time; $X$ and $Y$ mean the horizontal and vertical coordinates recorded by the GNSS locators; $\Delta t$ is 1s, matching the GNSS locator's frequency. It is noted that data for vessel No. 8 in Experiment 1 were missing from 14:07 to 14:23.

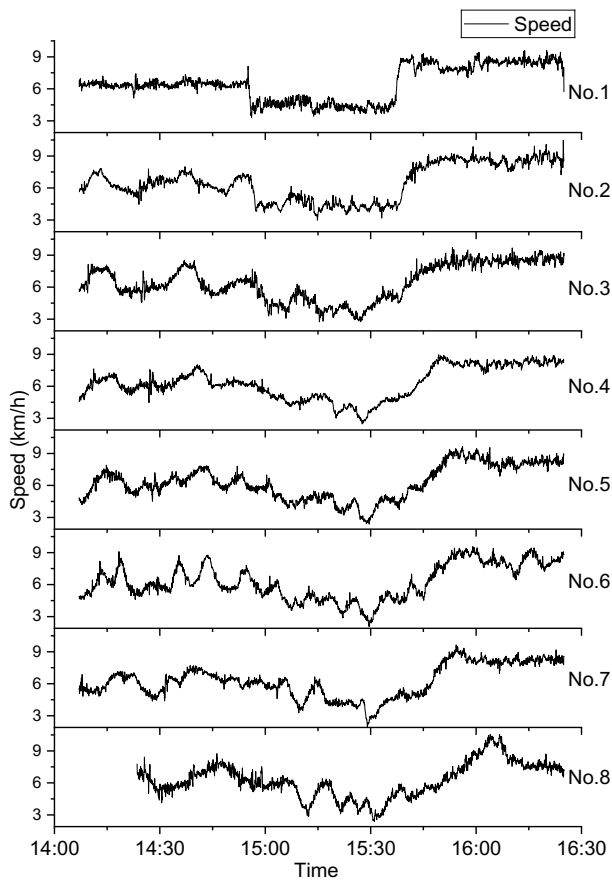

(a) Speed data of Experiment 1

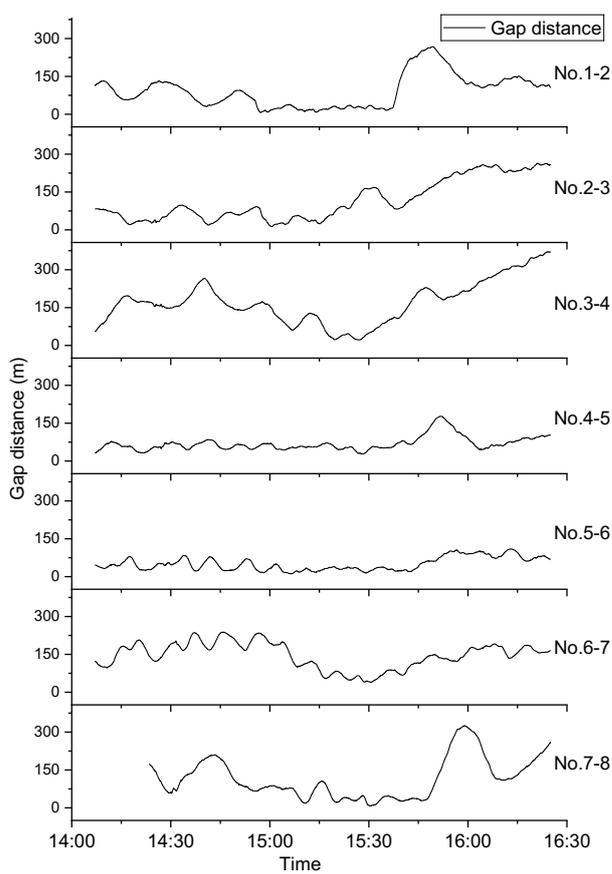

(b) Gap distance data of Experiment 1

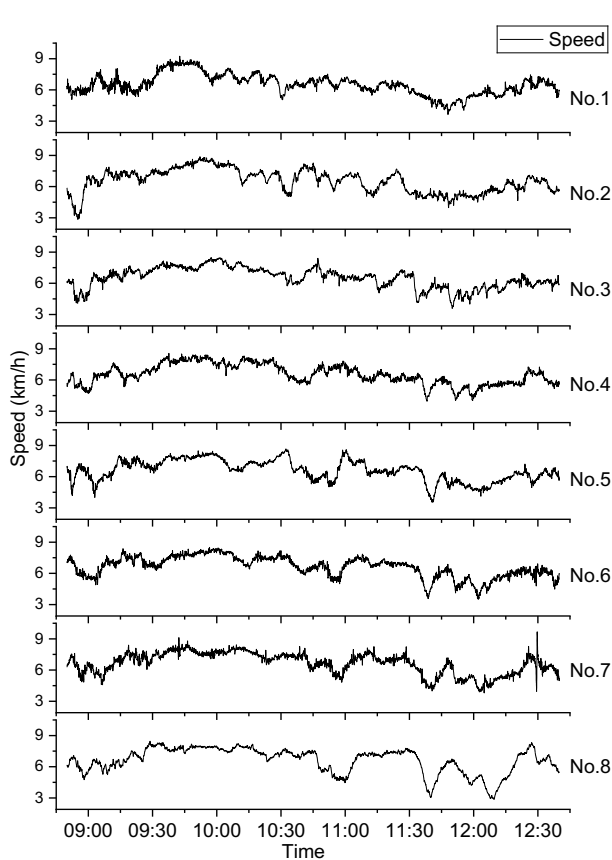

(c) Speed data of Experiment 2

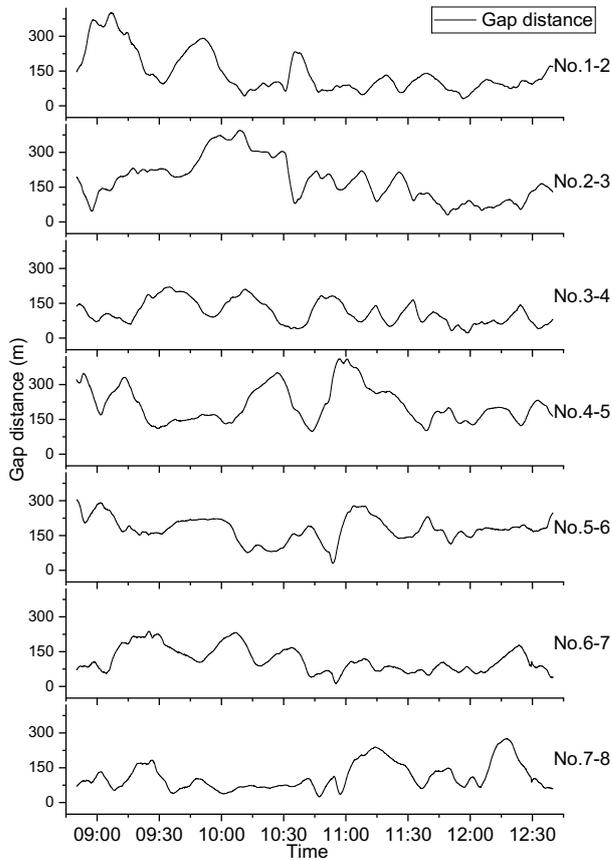

(d) Gap distance data of Experiment 2

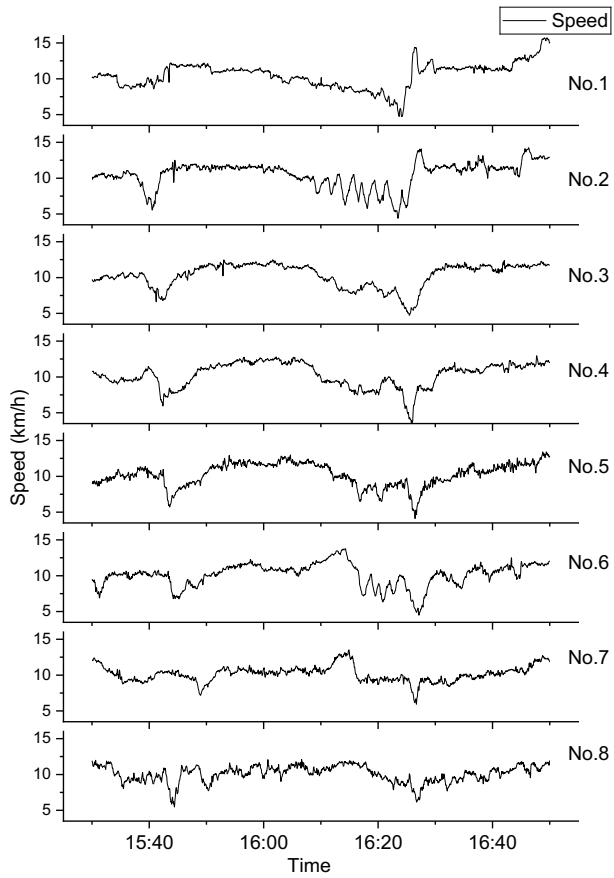 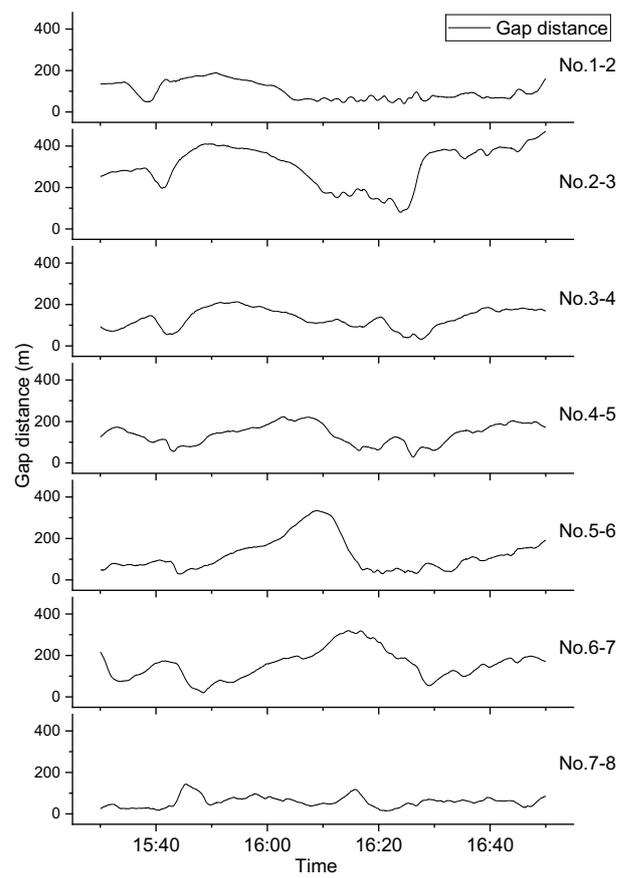

(e) Speed data of Experiment 3        (f) Gap distance data of Experiment 3

**Fig.6.** Speeds and gap distances of the experimental vessels.

## 3. Characteristic analysis

### 3.1. Video surveillance data

This section analyzes key parameters in vessel traffic flow, including flow rate, speed, and density, using video surveillance data. Furthermore, it determines the reference economic speed for vessels operating in inland restricted waterways based on these observations.

### 3.1.1. Parameter distribution

Fig. 7 depicts the distribution characteristics of flow rate, speed, and density based on video surveillance data. Table 4 provides the 15th percentile, median, 85th percentile, and average values for these parameters. The data reveals that vessel speeds are predominantly between 9 and 12 km/h. The average and median vessel speeds are nearly identical, suggesting a relatively symmetrical distribution. Flow rates and densities are primarily concentrated in the ranges of 20-40 vessels per hour and 1.5-4.5 vessels per kilometer, respectively, showing consistent frequency distributions. The generally low vessel density indicates favorable navigation conditions. Furthermore, the density distribution remains relatively concentrated, which suggests that the observed vessel traffic flow is continuous.

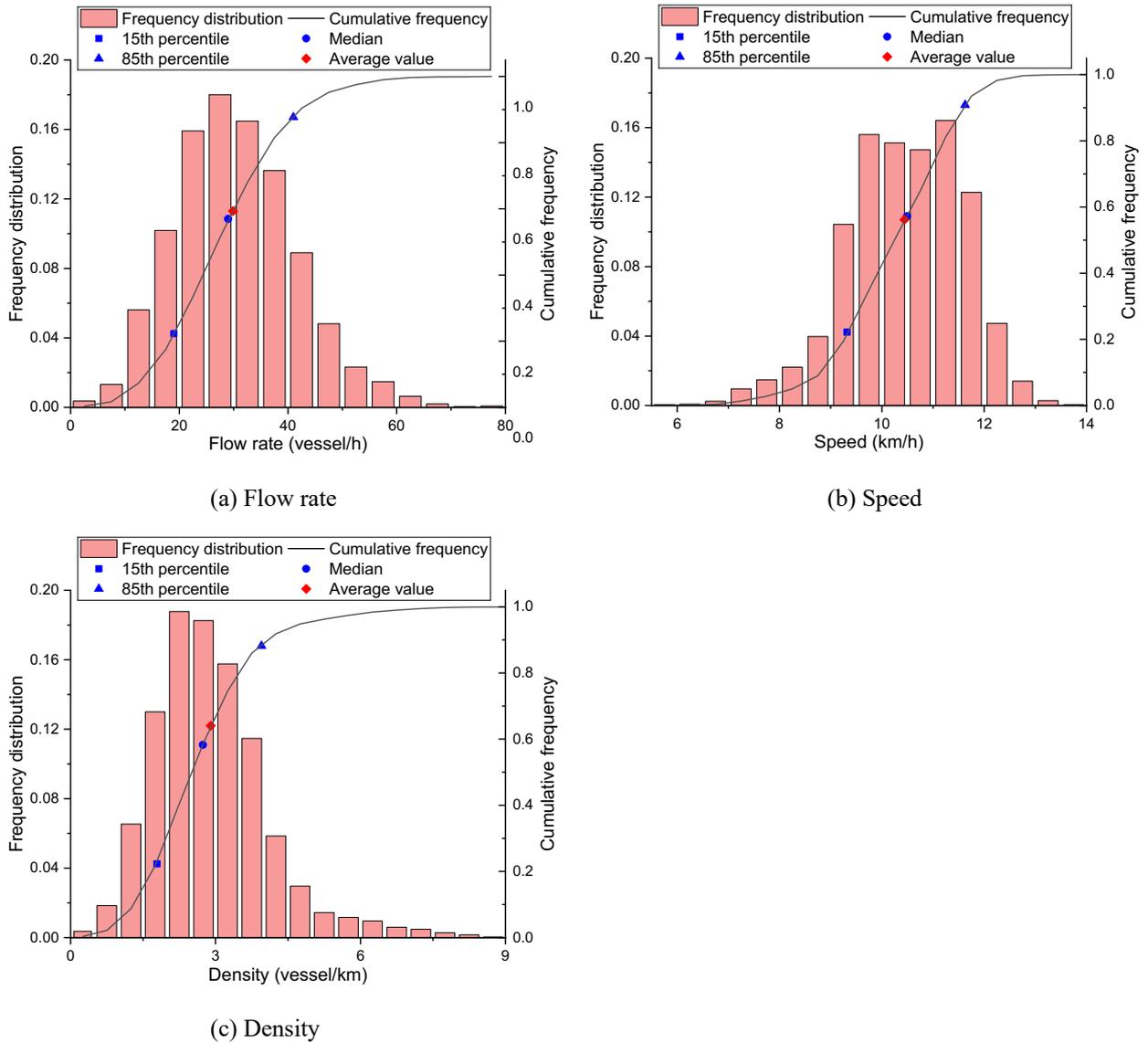

(a) Flow rate

(b) Speed

(c) Density

**Fig.7.** Parameter distribution of video surveillance data.

**Table 4**
The 15th percentile, median, 85th percentile, and average values for flow rate, speed, and density.

| Parameter | 15th percentile | Median | 85th percentile | Average value |
|---|---|---|---|---|
| Flow rate (vessel/h) | 19 | 29 | 41 | 29.89 |
| Speed (km/h) | 9.32 | 10.49 | 11.63 | 10.44 |
| Density (vessel/km) | 1.79 | 2.74 | 3.95 | 2.90 |

### 3.1.2. Economic speed

In most traffic scenarios, travelers aim to minimize travel time while ensuring safety to achieve the ultimate experience. For example, in road traffic, vehicles typically operate at the maximum safe speed allowed by the roads' speed limit, known as the free-flow speed. In contrast, inland vessels, even under safe free-flow conditions, often travel at reduced speeds. This is due to the high fuel costs associated with these vessels, which often outweigh the time costs. Consequently, the primary objective for these vessels is cost optimization rather than speed optimization. Vessel navigators must consider all related costs to make well-informed decisions.

To achieve cost optimization, vessel operators usually adopt an economic speed, which is defined as the speed that results in the lowest overall operating cost (Díaz-Secades et al., 2022; Hua et al., 2024). There is no standardized value for this speed; it is typically determined based on the navigators' experience. Economic speed can be considered analogous to the free-flow speed for vessels. Since video surveillance data captures vessel navigation characteristics under free-flow conditions, it is possible to derive the economic speed values for inland waterway vessels from this data.

Fig. 8 displays the speed distribution for vessels with varying loading conditions and lengths as recorded by video surveillance. The results indicate that vessel length has minimal impact on speed, suggesting that vessel size can be disregarded when determining economic speed. For vessels with different loading conditions, empty vessels generally travel at speeds between 11 km/h and 13 km/h, while loaded vessels typically travel between 8 km/h and 10 km/h. Therefore, using the median speed of vessels as an estimate for economic speed (illustrated by the line inside the box in Fig. 8(a)), it can be inferred that in inland restricted waterways of Yangtze River delta, the economic speed for empty vessels is approximately 12 km/h, whereas for loaded vessels, it is about 9 km/h.

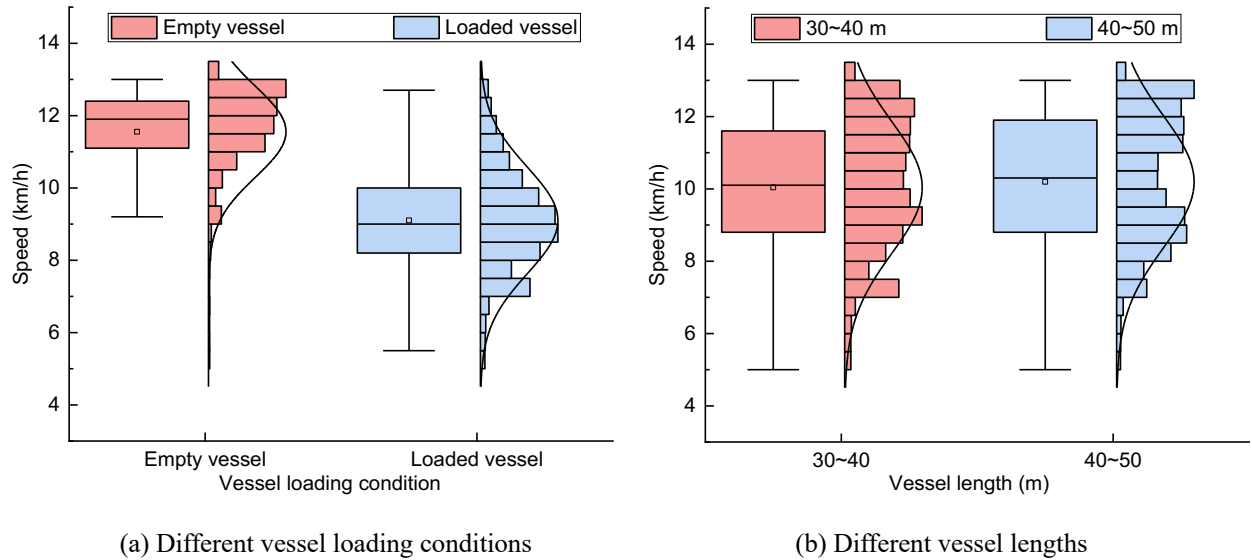

(a) Different vessel loading conditions                      (b) Different vessel lengths

**Fig.8.** The distribution of vessel speeds across varying loading conditions and vessel lengths.

### 3.2. Vessel-following experiment data

This section presents the distribution of vessel speeds and gap distances derived from vessel-following experiment data. It offers recommended values for the minimum vessel speed and minimum gap distance in inland restricted waterways. Furthermore, it examines the relationship between vessel speed and gap distance.

#### 3.2.1. Parameter distribution

Figs. 9 and 10 depict the frequency distribution and cumulative frequency curves for parameters from the vessel-following experiments, focusing on vessel speed and gap distance. Fig. 9(a) shows that the speed distribution for loaded vessels generally ranges from 3 to 9 km/h, while empty vessels typically travel between 7 and 13 km/h. In comparison to Fig. 7(b), it is evident that the average speed in the experiments is lower than that observed in video surveillance data. This difference is attributed to the high vessel density in the experiment, which creates non-free-flow conditions and results in reduced speeds. Fig. 9(b) indicates that the gap distance between vessels primarily ranges from 50 to 250 meters in the experimental data. This required gap distance limits vessels from achieving free-flow speeds.

Fig. 10 demonstrates that, even though the vessel was in a non-free-flow condition during the experiment, its speed remained above zero. This suggests a critical minimum speed threshold necessary for maintaining rudder effectiveness and proper vessel control (Liao et al., 2019). This threshold is affected by vessel performance, navigation environment, and operational conditions (Liu et al., 2017a; Tsevdou et al., 2019). Falling below this minimum speed can lead to loss of rudder control, requiring significant time and fuel to regain operational status. Fig. 10 recommends a minimum speed of 2.65 km/h and a corresponding minimum gap distance of 12 meters in inland restricted waterways. Experimental data reveal that only 0.1% of the recorded values fall below these recommended thresholds.

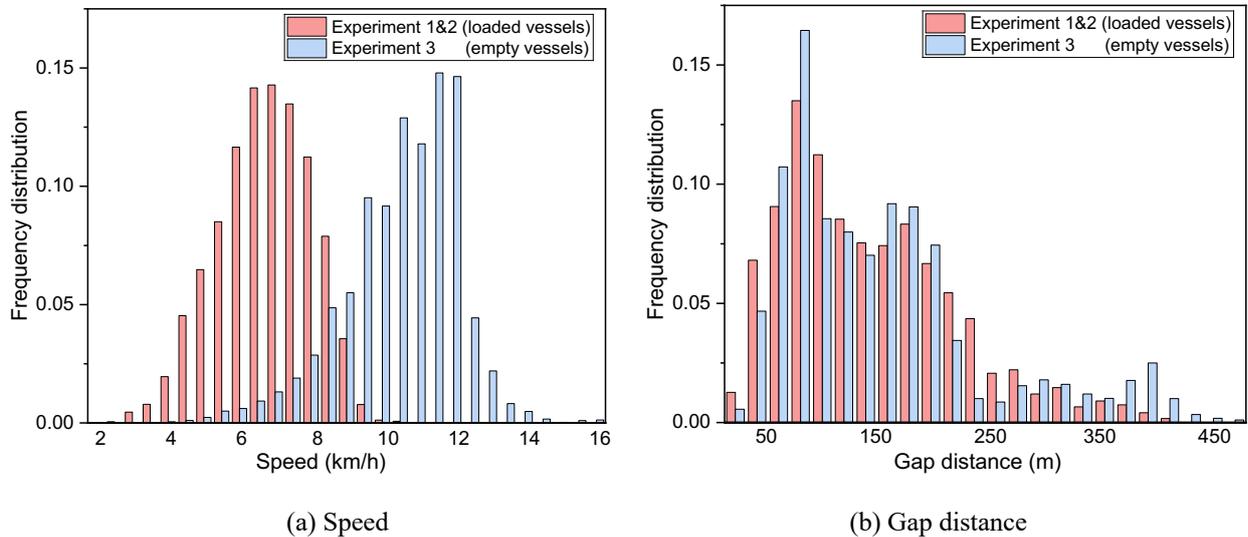

(a) Speed

(b) Gap distance

**Fig.9.** Parameter frequency distribution of vessel-following experiment data.

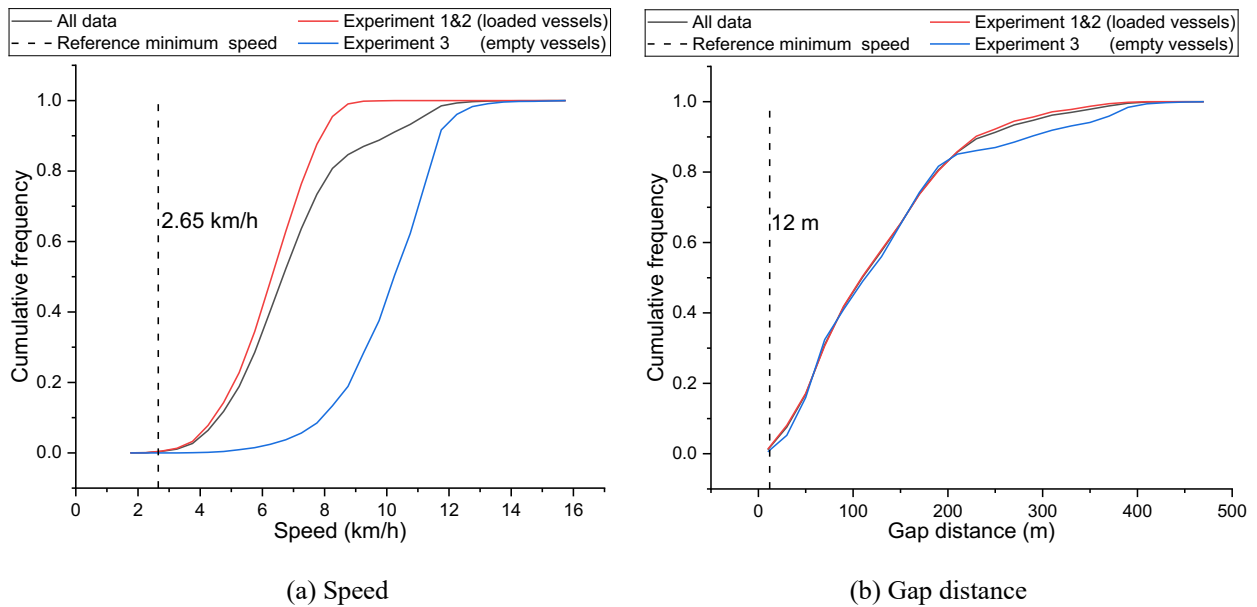

(a) Speed

(b) Gap distance

**Fig.10.** Parameter cumulative frequency curve of vessel-following experiment data.

### 3.2.2. Relationship between speed and gap distance

In vehicle traffic flow, various car-following models indicate that a vehicle aiming to travel at higher speeds must maintain a greater distance from the vehicle ahead (Chandler et al., 1958; Ni, 2016; Treiber et al., 2000). A similar phenomenon is observed in vessel-following experiments. For example, as shown in Fig. 6 (a) - (b), after 15:40, when the leading vessel begins to accelerate, the subsequent seven vessels also gradually increase their speeds. The gap distance between these vessels demonstrates a clear upward trend as vessel speed increases, suggesting a potential relationship between vessel speed and gap distance.

To investigate this relationship, we compiled a dataset correlating the speeds of the 2nd to 8th vessels in each experiment with their respective gap distances from the preceding vessel. Gap distances were categorized into 5-meter intervals, and the average vessel speed within each interval was calculated. The scatter plot of vessel speed versus gap distance is presented in Fig. 11. We employed four common function models—logarithmic, exponential, linear, and power functions—to fit the scatter plot. The fitting functions and their coefficients of determination ($R^2$) are summarized in Table 5. The calculation method for $R^2$ is as follows:

$$R^2 = \frac{\sum_{j=1}^{n}\left(\hat{x}_j - \overline{x}\right)^2}{\sum_{j=1}^{n}\left(x_j - \overline{x}\right)^2} \tag{3}$$

$$\overline{x} = \frac{1}{n}\sum_{j=1}^{n} x_j \tag{4}$$

where $n$ represents the number of data points; $j$ denotes the sequence index of the data; $x_j$ is the true value of the data at index $j$; $\hat{x}_j$ is the estimated value of the data at index $j$.

The analysis reveals a positive correlation between vessel speed and gap distance, with logarithmic and power function models providing the best fit. Due to its effectiveness and accuracy, the logarithmic function model is recommended for fitting. Additionally, a comparison between loaded vessels (Experiments 1 & 2) and empty vessels (Experiment 3) indicates that empty vessels travel at higher speeds for the same gap distance. Specifically, empty vessels are approximately 3-4 km/h faster at equivalent gap distances, as evidenced by the statistical results in Fig. 9 (a) and Table 5.

Understanding the relationship between speed and gap distance for vessels can aid in determining safe and efficient sailing speeds under non-free-flow conditions, supporting intelligent vessel operation. This insight is also valuable for vessel traffic modeling and simulation.

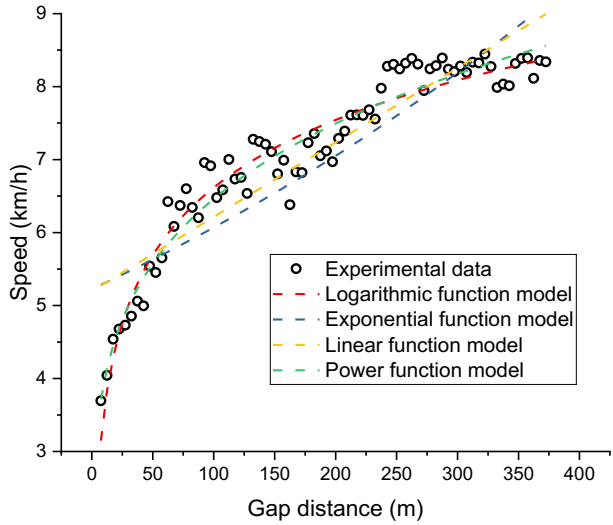

(a) Experiment 1

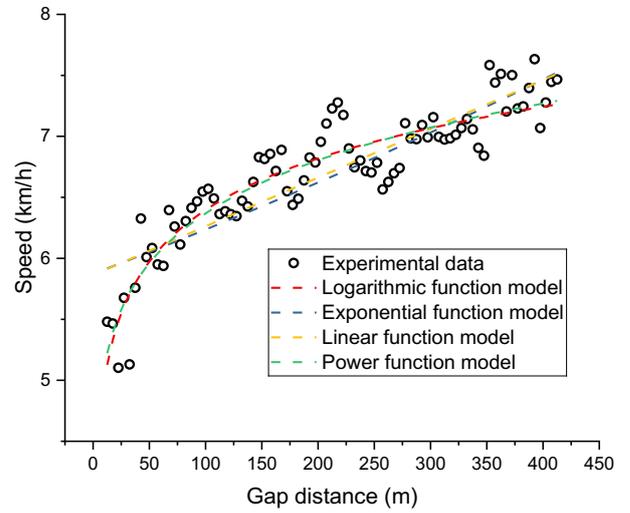

(b) Experiment 2

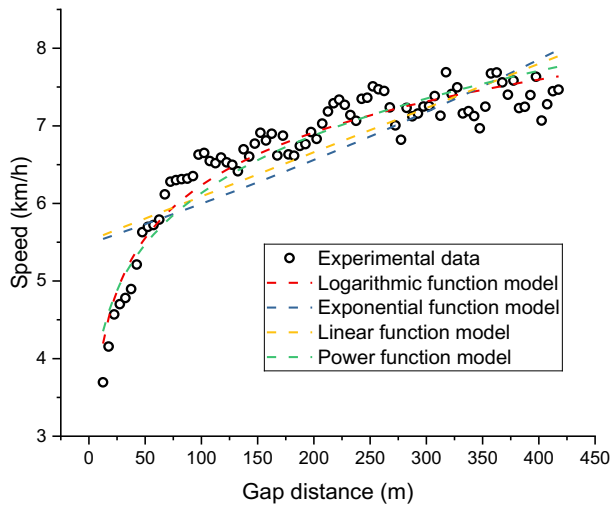

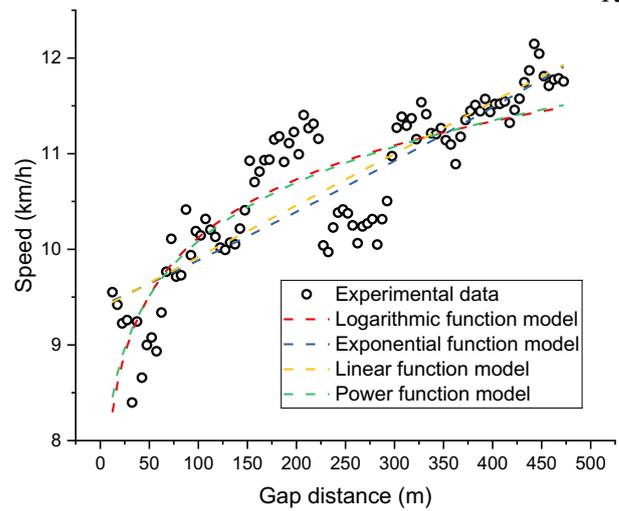

(c) Loaded vessels (Experiment 1&2)

(d) Empty vessels (Experiment 3)

**Fig.11.** Relationship between vessel speed and gap distance.

**Table 5**

Fitting functions for the relationship between vessel speed (km/h) and gap distance (m).

| Data classification | Model | Function | $R^2$ |
|---|---|---|---|
| Experiment 1 | Logarithmic function | $v = 1.3382\ln(g) + 0.4536$ | 0.94 |
| | Exponential function | $v = 5.224e^{0.0015g}$ | 0.79 |
| | Linear function | $v = 0.0102g + 5.1955$ | 0.83 |
| | Power function | $v = 2.4139g^{0.2138}$ | 0.94 |
| Experiment 2 | Logarithmic function | $v = 0.6102\ln(g) + 3.5863$ | 0.84 |
| | Exponential function | $v = 5.8728e^{0.0006g}$ | 0.75 |
| | Linear function | $v = 0.004g + 5.8622$ | 0.76 |
| | Power function | $v = 4.1042g^{0.0954}$ | 0.85 |
| Loaded vessels (Experiment 1&2) | Logarithmic function | $v = 0.9816\ln(g) + 1.714$ | 0.92 |
| | Exponential function | $v = 5.4803e^{0.0009g}$ | 0.64 |
| | Linear function | $v = 0.0057g + 5.5197$ | 0.67 |
| | Power function | $v = 2.8704g^{0.1649}$ | 0.89 |
| Empty vessels (Experiment 3) | Logarithmic function | $v = 0.8776\ln(g) + 6.0802$ | 0.73 |
| | Exponential function | $v = 9.4029e^{0.0005g}$ | 0.73 |
| | Linear function | $v = 0.0054g + 9.3791$ | 0.73 |
| | Power function | $v = 6.8223g^{0.0849}$ | 0.74 |

## 4. Fundamental diagram model

### 4.1. Traditional fundamental diagram model

The fundamental diagram model of traffic flow characterizes the functional relationships between three primary parameters of traffic flow under steady-state conditions: flow rate, speed, and density. Since flow rate is defined as the product of speed and density, the fundamental diagram is frequently expressed as the relationship between speed and density. Notable fundamental diagram models used in vehicle traffic flow include the Greenshields model (Greenshields et al., 1935), the Greenberg model (Greenberg, 1959), and the Underwood model (Underwood, 1961). The formulations of these models are as follows:

$$\text{Greenshields model}: v = -C_1 k + C_2 \tag{5}$$

$$\text{Greenberg model}: v = -C_3 \ln(k) + C_4 \tag{6}$$

$$\text{Underwood model}: v = C_5 e^{-C_6 k} \tag{7}$$

where $C_1 \sim C_6$ are positive parameters. The Greenshields model indicates a linear relationship between the speed and density of traffic flow, the Greenberg model suggests a logarithmic relationship, and the Underwood model proposes an exponential relationship. Additionally, the signs of the parameters in all three models reveal a negative correlation between speed and density of traffic flow.

In traffic flow fundamental diagram models, speed represents the spatial average speed. During the preprocessing of video surveillance data, we have obtained hourly vessel density and average speed. For the vessel-following experiment data, we use the following equation to calculate the density and average speed of vessel traffic flow at each moment.

$$\bar{v}_j(t) = \frac{8}{\sum\limits_{i=1}^{8} 1\big/ v_{i,j}(t)} \, (j = 1, 2, 3.) \tag{8}$$

$$k_j(t) = \frac{7}{\sum\limits_{i=2}^{8} \big(g_{i,j}(t) + L_{i,j}\big)} \, (j = 1, 2, 3.) \tag{9}$$

After determining the density and average speed, we correlated these parameters to create a comprehensive dataset. Densities were categorized into 0.2-vessels/km intervals, and the average speed within each interval was calculated. Fig. 12 illustrates the scatter distribution of the processed data. We applied three fundamental diagram models of traffic flow to fit these data, with the results presented in Table 6. The fitting results indicate that all three models accurately represent the data from the vessel-following experiments. However, the Greenberg model showed poor performance with video surveillance data, which is attributed to the low vessel density in the video data. The Greenberg model is known to be unsuitable for low-density traffic scenarios (Ni, 2016).

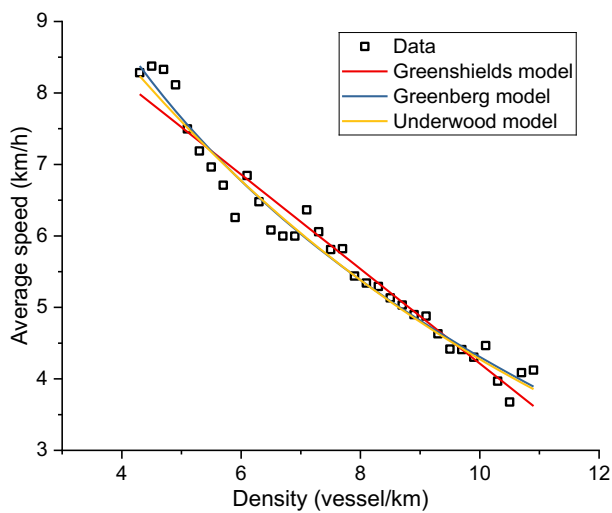

(a) Experiment 1

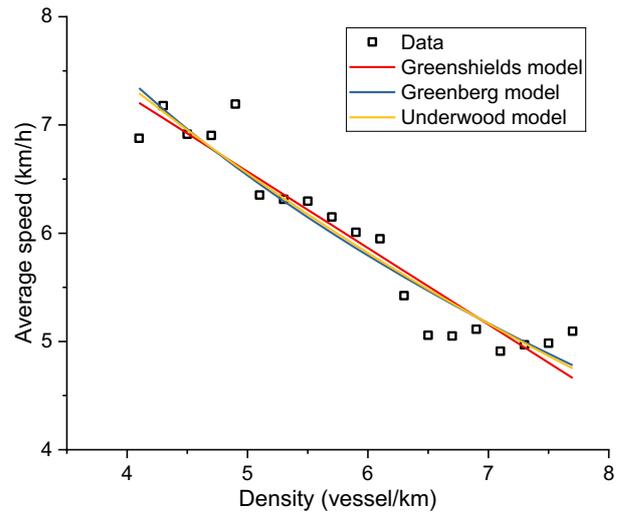

(b) Experiment 2

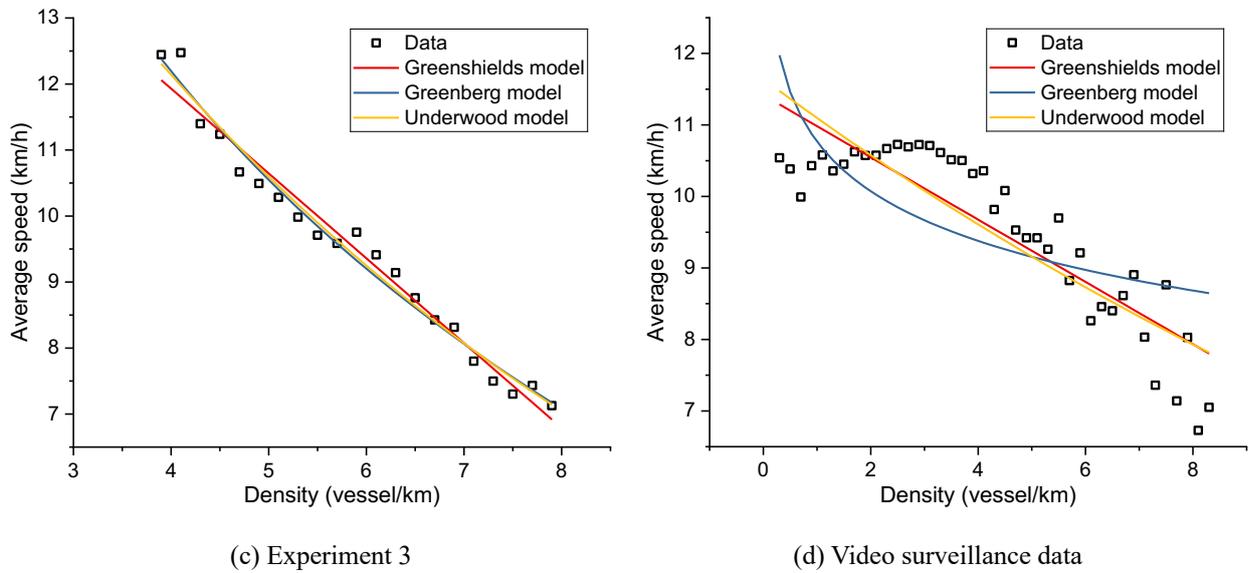

(c) Experiment 3          (d) Video surveillance data

**Fig.12.** Relationship between density and average speed.

**Table 6**
Fitting functions for the relationship between density (vessel/km) and average speed (km/h).

| Data classification | Model | Function | $R^2$ |
|---|---|---|---|
| Experiment 1 | Greenshields | $v = -0.6611k + 10.826$ | 0.95 |
| | Greenberg | $v = -4.8222\ln(k) + 15.411$ | 0.97 |
| | Underwood | $v = 13.504e^{-0.115k}$ | 0.97 |
| Experiment 2 | Greenshields | $v = -0.706k + 10.099$ | 0.91 |
| | Greenberg | $v = -4.0612\ln(k) + 13.07$ | 0.91 |
| | Underwood | $v = 11.88e^{-0.119k}$ | 0.91 |
| Experiment 3 | Greenshields | $v = -1.2854k + 17.072$ | 0.97 |
| | Greenberg | $v = -7.377\ln(k) + 22.42$ | 0.98 |
| | Underwood | $v = 20.915e^{-0.136k}$ | 0.98 |
| Video surveillance data | Greenshields | $v = -0.4355k + 11.418$ | 0.78 |
| | Greenberg | $v = -1.0022\ln(k) + 10.768$ | 0.48 |
| | Underwood | $v = 11.643e^{-0.048k}$ | 0.74 |

In traditional fundamental diagram models, three primary sets of characteristic parameter correspondences are identified, as outlined in Table 7. However, it is important to recognize that due to the presence of a vessel's minimum speed, the use of $k_j$ is not suitable for vessel traffic flow. Instead, maximum density ($k_{max}$) and minimum speed ($v_{min}$) should be used to define a new set of characteristic parameters. Consequently, $k = k_{max}$ and $v = v_{min}$ are employed instead of $k = k_j$ and $v = 0$.. By employing an appropriate fundamental diagram model to characterize traffic flow, characteristic parameter values can be extracted to represent various aspects of vessel traffic.

**Table 7**
Corresponding relationships of parameters in traffic flow fundamental diagram models.

| | Density ($k$) | Speed ($v$) | Flow rate ($q$) |
|---|---|---|---|
| 1 | 0 | $v_F$ (free-flow speed) | 0 |
| 2 | $k_j$ (jamming density) | 0 | 0 |
| 3 | $k_m$ (optimal density) | $v_m$ (optimal speed) | $q_m$ (maximum flow rate) |

In analyzing overall vessel traffic characteristics, it is beneficial to integrate and average data from both empty and loaded vessels, as inland shipping typically involves a combination of these conditions. By processing both video surveillance data and vessel-following experiment data collectively, the fundamental diagram of vessel traffic flow can be derived, as demonstrated in Fig.13. The characteristic parameter values are computed and summarized in Table 8.

While traditional traffic flow fundamental diagram models, such as the Greenshields model and the Underwood model, offer valuable insights into the relationships between flow rate, speed, and density in vessel traffic, some issues have been observed. For instance, there is a notable discrepancy in the speed-density relationship when vessel density is below 4 vessels per hour compared to when it exceeds 4 vessels per hour.

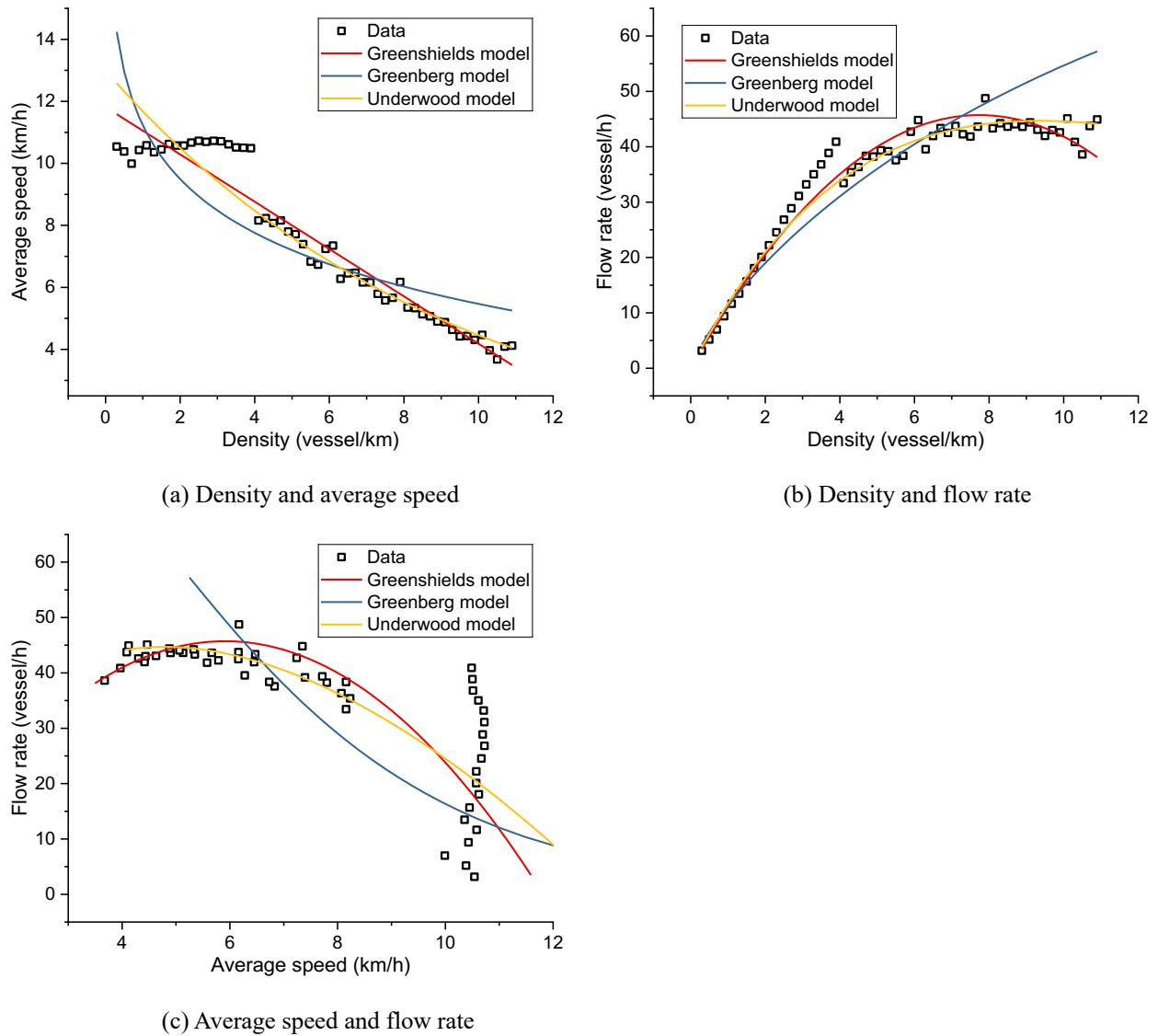

(a) Density and average speed

(b) Density and flow rate

(c) Average speed and flow rate

**Fig.13.** The results of applying traditional fundamental diagram models to vessel traffic flow data.

**Table 8**

Characteristic parameter values of vessel traffic flow and the overall fitting functions for the relationship between density (vessel/km) and average speed (km/h), based on traditional fundamental diagram models.

| Model | Function | $R^2$ | $v_f$ (km/h) | $v_m$ (km/h) | $k_m$ (vessel/km) | $q_m$ (vessel/h) | $k_{max}$ (vessel/km) |
|---|---|---|---|---|---|---|---|
| Greenshields | $v = -0.7634k + 11.817$ | 0.93 | 11.817 | 5.909 | 7.740 | 45.730 | 12.008 |
| Greenberg | $v = -2.502\ln(k) + 11.227$ | 0.71 | - | 2.650 | 30.834 | 81.709 | 30.834 |
| Underwood | $v = 12.999e^{-0.107k}$ | 0.90 | 12.999 | 4.782 | 9.346 | 44.692 | 14.863 |

### 4.2. Vessel traffic flow fundamental diagram model

[Fig. 13](#) demonstrates that traditional fundamental diagram models for traffic flow depict continuous relationships among traffic parameters. However, data on vessel traffic flow reveal a piecewise functional relationship. To address this discrepancy, the study proposes piecewise fundamental diagram models tailored for vessel traffic flow. According to these models, when $k < k_1$, the vessel traffic is in a free-flow condition, allowing vessels to travel at their economic speed, where $k_1$ represents the critical threshold between free-flow and non-free-flow conditions. Conversely, when $k > k_1$, the traffic flow shifts to a non-free-flow condition, causing vessels to deviate from their economic speed. In this regime, vessel speed experiences a noticeable and abrupt decline as density increases, a pattern observed in vehicle traffic flow as well ([Ni, 2016](#)). The structure of the piecewise models is outlined as follows:

$$\text{Piecewise model 1: } v = \begin{cases} v_f, & \text{if } k \le k_1 \text{ (free-flow condition)} \\ -C_1 k + C_2, & \text{if } k > k_1 \text{ (non-free-flow condition)} \end{cases} \tag{10}$$

$$\text{Piecewise model 2: } v = \begin{cases} v_f, & \text{if } k \le k_1 \text{ (free-flow condition)} \\ -C_3 \ln(k) + C_4, & \text{if } k > k_1 \text{ (non-free-flow condition)} \end{cases} \tag{11}$$

$$\text{Piecewise model 3: } v = \begin{cases} v_f, & \text{if } k \le k_1 \text{ (free-flow condition)} \\ C_5 \mathrm{e}^{-C_6 k}, & \text{if } k > k_1 \text{ (non-free-flow condition)} \end{cases} \tag{12}$$

As outlined in Section 3.1.2, in inland restricted waterways, the economic speed for empty vessels averages approximately 12 km/h, while for loaded vessels, it is about 9 km/h. To provide a comprehensive analysis of vessel characteristics, the overall economic speed is calculated as the average of these two speeds, resulting in an assumed value of 10.5 km/h for $v_f$. Piecewise models were employed to analyze the relationships between flow rate, speed, and density, as illustrated in [Fig. 14](#) and [Table 9](#). The $R^2$ values for the piecewise model fits are 0.98, 0.98, and 0.99, respectively, indicating a superior fit compared to traditional fundamental diagram models. Furthermore, [Fig. 14](#) shows that the piecewise models' line types more accurately represent the collected data, with piecewise model 3 providing the best fit.

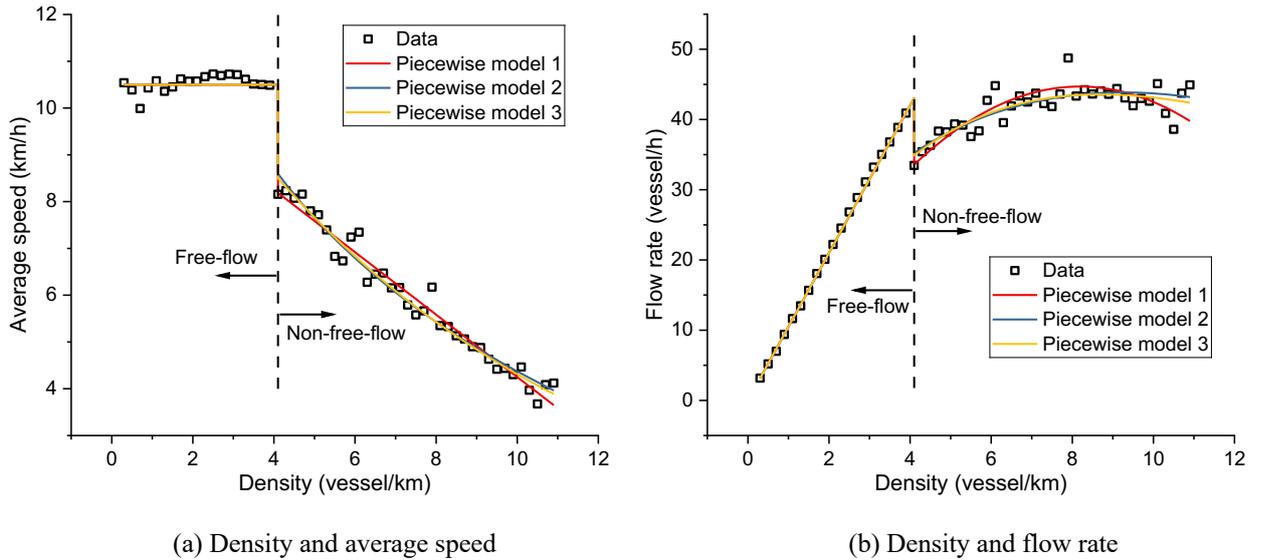

(a) Density and average speed          (b) Density and flow rate

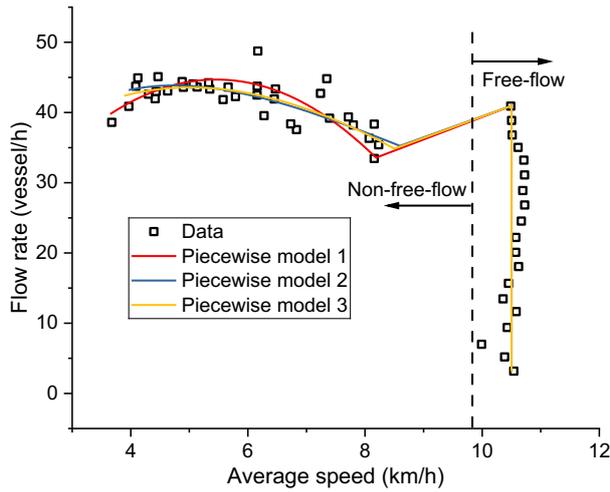

(c) Average speed and flow rate

**Fig.14.** The results of applying piecewise fundamental diagram models to vessel traffic flow data.

**Table 9**

Characteristic parameter values of vessel traffic flow and the overall fitting functions for the relationship between density (vessel/km) and average speed (km/h), based on the piecewise fundamental diagram models.

| Model | Function | $R^2$ | $v_f$ (km/h) | $v_m$ (km/h) | $k_m$ (vessel/km) | $q_m$ (vessel/h) | $k_{max}$ (vessel/km) |
|---|---|---|---|---|---|---|---|
| Piecewise model 1 | $v = \begin{cases} 10.5, \text{ if } k \leq 4 \\ -0.667k + 10.92, \text{ if } k > 4 \end{cases}$ | 0.98 | 10.5 | 5.462 | 8.184 | 44.700 | 12.398 |
| Piecewise model 2 | $v = \begin{cases} 10.5, \text{ if } k \leq 4 \\ -4.74\ln(k) + 15.28, \text{ if } k > 4 \end{cases}$ | 0.98 | 10.5 | 4.737 | 9.257 | 43.855 | 14.383 |
| Piecewise model 3 | $v = \begin{cases} 10.5, \text{ if } k \leq 4 \\ 13.62e^{-0.115k}, \text{ if } k > 4 \end{cases}$ | 0.99 | 10.5 | 5.011 | 8.696 | 43.573 | 14.235 |

Analyzing fundamental diagram models offers significant insights into vessel traffic flow. For instance, the theoretical capacity of a single waterway is indicated by the maximum flow rate ($q_m$), which is approximately 44 vessels per hour for inland restricted waterways. Fig. 14 shows that as vessel density approaches the critical threshold of 4 vessels per kilometer—the dividing line between free-flow and non-free-flow—the traffic flow rate nears 42 vessels per hour, close to this theoretical capacity. Beyond this density, the average vessel speed decreases and the flow rate increases only marginally. Therefore, to minimize delays and improve economic efficiency in inland restricted waterways, it is crucial to maintain vessel density below 4 vessels per kilometer.

## 5. Traffic state recognition

Severe vessel congestion can occur in certain sections and periods of inland waterways. Identifying the degree of congestion is crucial for efficient control and effective guidance of vessel traffic, which ensures smooth and orderly flow in these waterways. While inland waterway traffic conditions are generally categorized into free-flow and non-free flow, using these categories alone to assess congestion is overly simplistic and necessitates a more detailed classification.

K-means clustering is a widely used algorithm for clustering analysis (Bernardini, 2024; Pei et al., 2023). Its origins can be traced back to MacQueen (1965). The K-means clustering algorithm begins with a predetermined number of clusters and iteratively adjusts the allocation of data points to these clusters. The goal is to minimize the variance within each cluster while maximizing the variance between clusters. Specifically, the objective function aims to minimize the sum of squared distances between all data points and their respective cluster centers:

$$f(y) = \operatorname{argmin} \sum_{i=1}^{K} \sum_{j=1}^{n_i} \left\| y_j - c_i \right\|^2 \tag{13}$$

$$c_i = \frac{1}{n_i} \sum_{j=1}^{n_i} x_i, \, (i = 1, ..., K) \tag{14}$$

where $f(x)$ represents the variable value that minimizes the sum of Euclidean distance errors between all data points and their corresponding cluster centers, which is used to identify data points across all clusters. $x_j$ denotes a data point within cluster $i$; $K$ is the total number of clusters, and $n_i$ is the number of data points in cluster $i$. $c_i$ refers to the center of the $i$-th cluster.

After applying the K-means clustering algorithm, the silhouette coefficient can be used as an evaluation index for clustering, with the calculation formula provided below:

$$s = \frac{b - a}{\max(a, b)} \tag{15}$$

where $s$ represents the silhouette coefficient, which ranges from -1 to 1. A value closer to 1 indicates a better clustering effect. $a$ denotes the average distance between a data point and all other data points within the same cluster, while $b$ denotes the average distance between a data point and all other data points in the nearest cluster.

This study uses the k-means clustering algorithm to analyze vessel traffic speed data, which includes all samples from both video surveillance and vessel-following experiments. Fig. 15 illustrates the silhouette coefficients for clustering results with varying numbers of clusters. It is evident that the silhouette coefficient is highest when the number of clusters is 4, suggesting that this clustering configuration yields the most reasonable results. Therefore, the study classifies vessel traffic states into four distinct types.

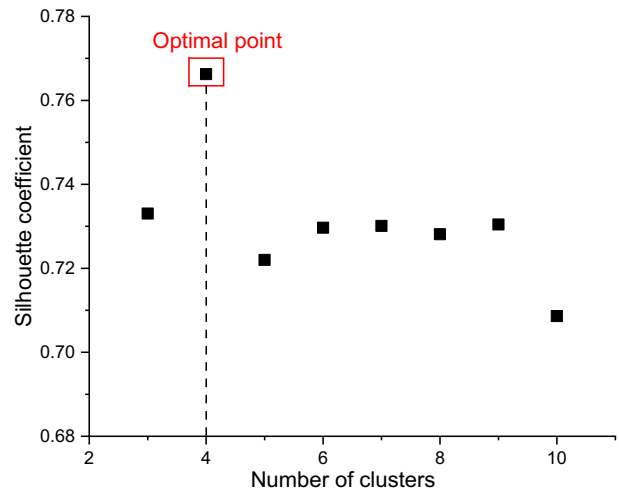

**Fig.15.** Silhouette coefficients for clustering results with varying numbers of clusters.

Table 10 presents the average speed ranges for different traffic states on waterways. In practice, continuously monitoring vessel speed along an entire route can be quite challenging. Instead, it is more practical to determine flow rate and density by counting the number of vessels passing fixed points. Since there is a known relationship between vessel traffic flow rate, density, and speed, average vessel speed can be estimated using real-time data on flow rate and density. In the density-flow rate distribution diagram, the slope of the line passing through the origin represents velocity. This speed can then be used to classify congestion states, as demonstrated in Fig. 16.

**Table 10**
Average speed ranges for different traffic states.

| Waterway traffic state | Smooth | Slow | Congested | Severely congested |
|---|---|---|---|---|
| Average speed range (km/h) | > 9.38 | (7.28, 9.38] | (5.67, 7.28] | ≤ 5.67 |

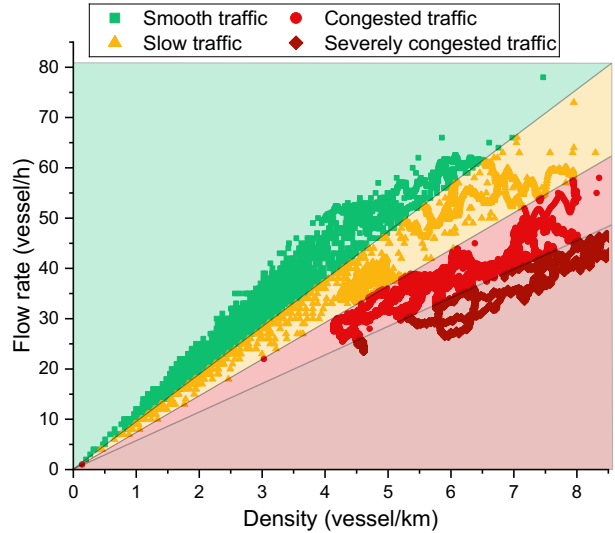

**Fig.16.** Results of vessel traffic flow state recognition in inland restricted waterways.

In inland waterway vessel navigation services, real-time data on flow rates and density are used to represent traffic conditions on electronic waterway maps. The routes are color-coded—green, yellow, red, and dark red—similar to the system used in vehicle navigation on city roads. This system enhances navigation services for crew members by providing more accurate and up-to-date information. This feature has been incorporated into the Jiangsu Inland Waterway Vessel Navigation mobile app. Fig. 17 illustrates how different waterway traffic states are displayed across various navigation interfaces within the app.

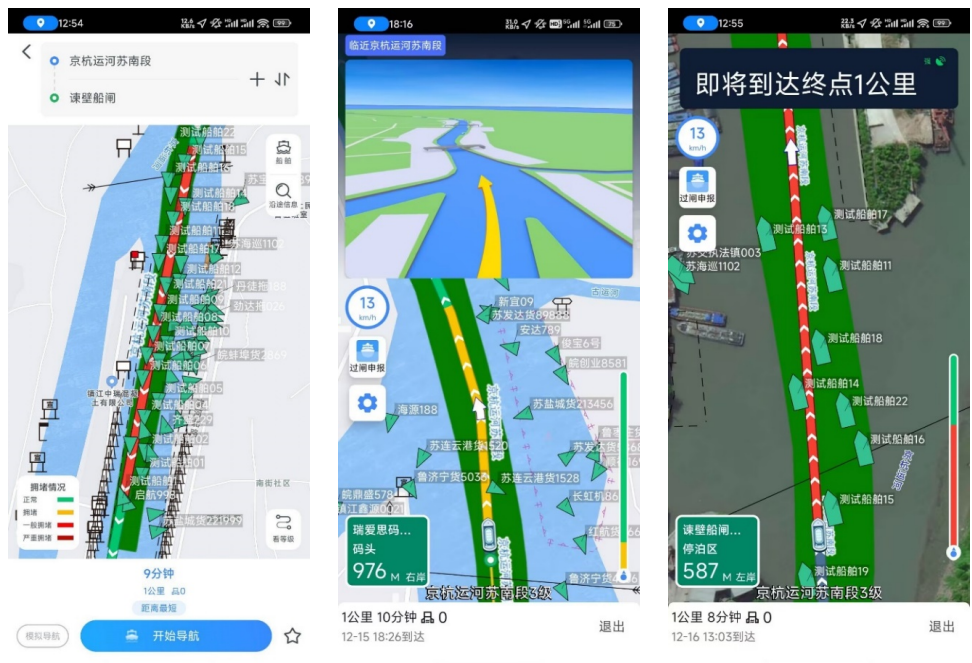

**Fig.17.** The interface of the Jiangsu inland waterway vessel navigation app.

# 6. Conclusions

This study investigates the characteristics of vessel traffic flow in inland restricted waterways using multi-source data, specifically video surveillance data and vessel-following experiment data. The study is structured as follows:

(1) Initially, the study analyzed macroscopic flow rate, speed, and density using video surveillance data. It was observed that in inland restricted waterways of Yangtze River delta, the economic speed for empty vessels averages around 12 km/h, while for loaded vessels, it is approximately 9 km/h.

(2) Next, the study explored microscopic aspects such as vessel speed and gap distance through vessel-following experiments. Findings indicated that the minimum vessel speed in inland restricted waterways is about 2.65 km/h, with a minimum gap distance of 12 meters. Various models—including logarithmic, exponential, linear, and power functions—were tested to describe the relationship between vessel speed and gap distance. The logarithmic model was found to be the most effective and accurate.

(3) Furthermore, the relationship between macroscopic speed, density, and flow rate was investigated. A piecewise vessel traffic flow fundamental diagram model was proposed, which better captured vessel traffic characteristics. The model suggests that the maximum flow rate is approximately 44 vessels per hour and recommends maintaining vessel density below 4 vessels per kilometer to minimize delays and enhance economic efficiency.

(4) Lastly, the K-means clustering algorithm was applied to categorize vessel traffic flow into four levels: smooth, slow, congested, and severely congested. The study identified average speed ranges for each level and show its application in inland waterway navigation.

The aim of this study is to provide theoretical support for advancing inland waterway transportation and to contribute to the development of a modern and comprehensive waterway transportation system. Future research will focus on expanding data sources, enhancing the reliability of analysis results, and further exploring vessel traffic flow characteristics in restricted inland waterways from additional perspectives.

## CRediT authorship contribution statement

**Wenzhang Yang**: Methodology; Investigation; Funding acquisition; Supervision; Formal analysis; Software; Writing - original draft; Writing - review & editing. **Peng Liao**: Conceptualization; Methodology; Investigation; Funding acquisition; Supervision; Formal analysis; Writing - original draft; Writing - review & editing. **Shangkun Jiang**: Investigation; Formal analysis; Software; Writing - original draft; Writing - review & editing. **Hao Wang**: Supervision; Writing - review & editing.

## Declaration of competing interest

The authors declare that they have no known competing financial interests or personal relationships that could have appeared to influence the work reported in this paper.

## Acknowledgements

This work was sponsored by the National Natural Science Foundation of China (No. 52172303), the SEU Innovation Capability Enhancement Plan for Doctoral Students (CXJH_SEU 24178), and the Postgraduate Research & Practice Innovation Program of Jiangsu Province (KYCX24_0451).